\magnification=\magstep 1
\overfullrule=0pt
\hfuzz=16pt
\voffset=0.0 true in
\vsize=8.8 true in
\baselineskip 20pt
\parskip 6pt
\hoffset=0.1 true in
\hsize=6.3 true in
\nopagenumbers
\pageno=1
\footline={\hfil -- {\folio} -- \hfil}

\centerline{{\bf Exact Solutions of Anisotropic
Diffusion-Limited}}

\centerline{{\bf Reactions with Coagulation and
Annihilation}}

\vskip 0.4in

\centerline{{\bf Vladimir Privman},\ \ {\bf Ant\'{o}nio
M.~R.~Cadilhe},\ \ and\ \ {\bf M.~Lawrence Glasser}}

\vskip 0.2in

\centerline{\sl Department of Physics, Clarkson University,
Potsdam, New York 13699--5820, USA}

\vskip 0.4in

\centerline{\bf ABSTRACT}

We report exact results for one-dimensional
reaction-diffusion models $A+A \to {\rm inert}$, $A+A \to
A$, and $A+B \to {\rm inert}$, where in the latter case
like particles coagulate on encounters and move as
clusters. Our study emphasized anisotropy of hopping rates;
no changes in universal properties were found, due to
anisotropy, in all three reactions. The method of solution
employed mapping onto a model of coagulating positive
integer charges. The dynamical rules were synchronous,
cellular-automaton type. All the asymptotic large-time
results for particle densities were consistent, in the
framework of universality, with other model results with
different dynamical rules, when available in the
literature.

\ 


\ 

\noindent {\bf KEY words:}\ \ reactions, anisotropic diffusion,
coagulation, synchronous dynamics

\ 

\noindent {\bf Running title:}\ \ Anisotropic diffusion-limited
reactions

\vfil\eject

\noindent {\bf I. INTRODUCTION}

\

Diffusion-Limited Reactions (DLR) involving aggregation and
annihilation processes are important in many physical,
chemical and biological phenomena [1-3] such as star
formation, polymerization, recombination of charge carriers
in semiconductors, soliton and antisoliton annihilation,
biologically competing species, etc. In this paper we
generalize and apply a recently introduced method [4-5], in
order to study by exact solution effects of anisotropy in
some common DLR in one dimension ($1D$), specifically,
$A+A\to A$, $A+A\to{\rm inert}$, and a two-species
annihilation model $A+B\to{\rm inert}$ in which like
particles coagulate irreversibly. The detailed definitions
will be given later.

Scaling approaches and other methods have yielded
[1-3,6-12] the upper critical dimension, $D_c$, for various
reactions. Typical values range from 2 to 4. For spatial
dimensions lower than $D_c$ the kinetics of these reactions
is fluctuation-dominated, and we cannot expect the rate
equation approach to be valid. Indeed the mean-field rate
equation approximation ignores effects of inhomogeneous
fluctuations. Fluctuation-dominated DLR have been subject to
numerous studies by other methods, notably, exact solutions
and asymptotic arguments [1,13-22] in $1D$. The $1D$
reactions have also found some experimental applications
[23-24]. These studies have assumed isotropic hopping
(reactant particle diffusion).

Recently Janowsky [25] concluded, based on numerical
results and phenomenological considerations for the two
species annihilation reaction $A+B\to{\rm inert}$, that
making the hopping fully directed would change the
universality class in $1D$. Specifically, the large-time
particle concentration (assuming equal densities of both
species), would scale according to $c (t) \sim t^{-1/3}$
instead of the isotropic-hopping power law $t^{-1/4}$. A
few exact and numerical results available in the literature
on anisotropic reactions involving only one species [26-27]
indicate that the power law is not changed. The model of
[25] assumed that like particles interact via hard-core
repulsion; this seems
to be an essential ingredient for observing the changeover
in the universality class.

In this work we report the exact solution for two-particle
annihilation with anisotropic hopping. We consider
discrete-time simultaneous-updating dynamics, also termed
synchronous dynamics, so that our models are
cellular-automaton type. The universality classes of
behavior at large times and large spatial scales are
expected not to depend on details of the model dynamics.
However, in order to achieve exact solvability we took
``sticky-particle'' rather than hard-core interactions: the
like particles coagulate on encounters and diffuse as
groups. Our exact calculations yield the $t^{-1/4}$ power
law, found earlier for similar ``sticky-particle''
reactions with isotropic hopping [28-29]. This result is
probably due to absence of hard-core interactions in our
model.

For unequal initial concentrations, the large-time behavior
changes, see [28-29]. The crossover between the two regimes
is derived analytically. Finally, we also obtain new exact
results for single-species two-particle aggregation and
annihilation reactions with anisotropy; see also [26]. We
find that anisotropy does not change the universality class of
kinetics of these reactions. A short version of this work was
reported in a letter-style publication [30].

This paper is organized as follows: in Section~II we define
and review some existing results for the models studied. In
Section~III our method for exact solution is introduced. In
section IV, results for the single-species models are
presented. Finally, Section~V is devoted to the two-species
model. Further scaling analysis of the two-species model
and the brief summarizing discussion are left to Section~VI.

\vfil\eject

\noindent {\bf II. REACTION-DIFFUSION MODELS}

\

In lattice DLR models with like particles it is usually
assumed that the particles hop independently, to the extent
allowed by their interactions, to their nearest neighbor
sites. Whenever two particles meet, they can both
annihilate which corresponds to the reaction $A+A\to {\rm
inert}$. If, however, only one particle of a pair
disappears we get the reaction $A+A\to A$ usually termed
``aggregation.''\ Such single-species reactions have the
upper critical dimension $D_c=2$\ [6-9]. For $D<2$ the
particle concentration at large times behaves according to
$c (t) \sim t^{-D/2}$. Specifically, the $1D$ kinetics
of these reactions is non-mean-field, with the typical
diffusional behavior $c (t) \sim t^{-1/2}$. This result
is not affected by short-range interactions between the
particles and is not sensitive to their initial
distribution as long as initial correlations are
sufficiently short-range.

Consider now the two-particle annihilation model, to be
termed the $AB$ model. Particles hop randomly to one of
their nearest-neighbor sites. Whenever two particles meet,
unlike species annihilate, $A+B\to {\rm inert}$. When like
species meet, some interaction must be assumed. The
simplest interaction is hard-core: diffusion attempts
leading to multiple occupancy of lattice sites are
discarded. For such reactions, assuming equal average $A$-
and $B$-concentrations and uniform initial conditions, the
upper critical dimension is $D_c=4$ [6-9]. The (equal)
particle concentrations scale according to $c (t) \sim
t^{-D/4}$. A surprising largely numerically-based recent
result of Janowsky [25] is the new exponent $\approx 1/3$,
replacing $1/4$, for anisotropic particle hopping in $1D$.
For unequal initial concentrations, the density of the
minority species is expected to decay faster than the
symmetric-case power-law; some specific results will be
referred to later.

In order to obtain a solvable model in $1D$, let us now
consider the $AB$ annihilation model with the ``sticky
particle'' interaction. Thus, like particles coagulate
irreversibly on encounters, for instance,

$$nA+mA\to (n+m)A \;\; , \eqno(2.1)$$

\noindent and the clusters thus formed then diffuse as
single entities with the single-particle hopping rates.
When unlike clusters meet at a lattice site, the outcome of
the reaction is

$$ nA+mB \to \cases{ (n-m)A \;\; & if $\;\;\; n>m$ \cr {\rm
inert} \;\; & if $\;\;\; n=m $ \cr (m-n)B \;\; & if $\;\;\;
n<m$ } \eqno(2.2)$$

\noindent Recent numerical results and scaling
considerations for these reactions [31] in $D=1,2,3$ seem
to suggest that they are mean-field in $D=2,3$. However, in
$1D$ the power-law exponent for the density is $1/4$,
obtained by exact solution [28-29] which also yielded a
faster power-law decay $\sim t^{-3/2}$ for the minority
species in case of unequal densities of $A$ and $B$.

As mentioned earlier, in order to obtain exact solutions
for our single-species and $AB$ models, we first solve
another model [4], of coagulating nonnegative integer
charges in $1D$\ [4-5], with anisotropic hopping. All our
models are defined with synchronous dynamics to be
described in detail in the next section, along with
detailed dynamical rules and their exact solution. From
the coagulating-charge model one can derive results for the
reactions $A+A \to A$ and $A+A \to {\rm inert}$ with
anisotropic hopping; see Section~IV. Some of our
expressions are new, while other are consistent with
results available in the literature. We next use the
approach of [28-29] to extend the coagulating-charge
solution to the ``sticky'' $AB$ model. We find that
hopping anisotropy does not change the density exponent in
$1D$:\ it remains $1/4$; see Section~V.

\vfil\eject

\noindent {\bf III. SYNCHRONOUS DYNAMICS OF COAGULATING
CHARGES}

\

Let us consider a one-dimensional lattice with unit
spacing. Following [4], we consider diffusion of
nonnegative charges on this lattice. Initially, at $t=0$,
we place positive unit charge at each site with probability
$p$ or zero charge with probability $1-p$. Furthermore,
we consider synchronous dynamics, i.e., charges at all
lattice sites hop simultaneously in each time step $t \to
t+1$, where $t=0,1,2,\ldots$. However, the probabilities of
hopping to the right, $r$, and to the left, $\ell=1-r$, are
not necessarily equal. Since this dynamics decouples the
even-odd and odd-even space-time sublattices, it suffices
to consider only those charges which are at the even
sublattice at $t=0$. Thus we only consider the lattice
sites $j=0,\pm 2,\pm 4, \ldots$ at times $t=0,2,4,\ldots$,
and lattice sites $j=\pm 1, \pm 3 , \pm 5, \ldots$ at times
$t=1,3,5, \ldots$.

One can view the diffusional hopping as taking place on a
directed square lattice with the time direction along the
``directed'' diagonal. This lattice is illustrated in
Figure~1; note that the charges arriving at site $j$ at
time $t$ can only come from the sites $j-1$ and $j+1$ at
time $t-1$. Finally, the ``interaction'' between the
charges is defined by the rule that all charge accumulated
at site $j$ at time $t$ coagulates. There can be 0, 1 or 2
such charges arriving from the two nearest neighbors of $j$
in the time step $t-1 \to t$, depending on the random
decisions regarding the directions of hopping from sites $j
\pm 1$ in this time step.

This model can also be viewed as diffusion-coagulation of
unit-charge ``particles'' $C$, where the coagulation is
represented by the reaction

$$nC+mC\to (n+m)C \;\; . \eqno(3.1)$$

\noindent Such reactions, without the limitation of
positive or integer charges, and with an added process of
feeding-in charge at each time step, with values drawn
randomly at each site from some fixed distribution, have
been considered as models of self-organized criticality and
coagulation [5,32-33]. These studies were limited to
isotropic hopping, i.e., $r=\ell={1 \over 2}$. Our interest
in these reactions is in that their dynamics can be mapped
[4,28-29] onto that of both the single-species
(Section~IV) and ``sticky'' $AB$\ (Section~V)
reaction-diffusion models introduced in Section~II.
However, before discussing and utilizing this mapping, let
us present the exact solution of the model of coagulating
charges with anisotropic hopping, following the ideas of
[4-5].

For each time $t$ and at each lattice site $j$ (of the
relevant sublattice) we define stochastic variables,

$$\tau_j (t)=\cases{1,&\ probability $r$ \cr 0,&\ probability
$\ell$} \eqno(3.2)$$

\noindent which represent the hopping direction decisions.
Then the stochastic equation of motion for the charges $q_j
(t)$, equal to the number of $C$ particles at site $j$ at time
$t$, is

$$q_n (t+1) = \tau_{n-1} (t) q_{n-1} (t) + \left[ 1-\tau_{n+1}
(t) \right] q_{n+1} (t) \;\; . \eqno(3.3)$$

The total number of $C$-particles, or the total charge, in
an interval of $k$ consecutive proper-parity-sublattice
sites, starting at site $j$ at time $t$, is given by

$$S_{k, j} (t)=\sum_{i=0}^{k-1} q_{j+2i} (t)=q_j
(t)+q_{j+2} (t)+\cdots+ +q_{i+2k-2} (t) \;\; . \eqno(3.4)$$

\noindent Due to conservation of charge in this process,
the equations of motion (3.3) yield the following relation,

\

\hang{}$ S_{k, n} (t+1)=\tau_{n-1} (t) q_{n-1} (t)$
\hfill\break $ \hbox{\hphantom{$S_{k, n} (t+1)$}} +q_{n+1}
(t)+\cdots+q_{n+2k-3} (t)$ \hfill\break $
\hbox{\hphantom{$S_{k, n} (t+1)$}} + \left[1-\tau_{n+2k-1}
(t)\right]q_{n+2k-1} (t) \;\;$. \hfill (3.5)\break

\noindent This result indicates that only the two random
decisions at the end points are involved in the dynamics of
charges in consecutive-site intervals. The exact
solvability of coagulating-charge models is based on this
property, as first observed in [5]. Other solution methods
were used in related interface-growth models [34] which
will not be discussed here.

Let us introduce the function

$$ I(s,m) = \delta_{s , m} \;\; , \eqno(3.6) $$

\noindent and averages,

$$ f_{k, m} (t) = \langle I \left( S_{k, j}(t),m \right)
\rangle \;\; . \eqno(3.7)$$

\noindent The average $\langle \cdots \rangle$ is over the
stochastic dynamics, i.e., over random choices of the
decision variables $\tau_i (t)$, as well as over the random
initial conditions. Since the latter are uniform, the
averages $f_{k, m} (t)$ in (3.7) do not depend on the
lattice site $j$. Functions other than the Kronecker delta
have been used for $I(s,m)$ in the literature [4-5,32-33].
With our choice (3.6), the resulting averages $f_{k, m}
(t)$ correspond to the probability to find exactly $m$
charge units in an interval of $k$ consecutive sites. For
instance, $f_{1, m} (t)$ is the density (fraction) of sites
with charge $m$.

Note that (3.5) essentially represents the following simple
rule: charge ``fed'' into an interval of $k$
sites comes from $k+1$ sites, at the preceding
time-variable value, with probability $rl$, from
$k-1$ sites with probability
$rl$, and from two possible groups of $k$ sites, with
probabilities $r^2$ and $\ell^2$; this is illustrated in
Figure~2. Furthermore, the variables $\tau_i (t)$ and $S_{k,
n} (t)$ are statistically independent because the latter
only depends on ``decision making'' variables $\tau_i$ at
earlier times. Therefore, the averages introduced in (3.7)
satisfy, for any function $I(s,m)$, the following equations
of motion,

$$ f_{k,m} (t+1) =r\ell\left[f_{k+1,m} (t)+f_{k-1,m}
(t)\right]+\left(r^2+\ell^2\right)f_{k,m}
(t) \;\; . \eqno(3.8)$$

Interestingly, the $m$-dependence is parametric in (3.8).
However, it does enter the initial conditions. It is also
convenient [4] to define $f_{0,m} (t) = I(0,m)$ in order to
extend the applicability of (3.8) to all $t \geq 0$. For
our specific choices, we have the following expressions.
Firstly, the initial conditions of placing charges 1 or 0
at each site, with respective probabilities $p$ and
$1-p$, correspond to

$$f_{k,m}(0)=\cases{p^m (1-p)^{k-m} {k\choose m}&$0\leq m\leq
k$\cr 0&\vphantom{$k\choose m$}$m>k$} \eqno(3.9)$$

\noindent where ${k\choose m}=k!/\left(m!(k-m)!\right)$.
Secondly, the boundary condition is that a null interval
cannot have charges, i.e.,

$$f_{0,m}(t\geq 0)= \delta_{0,m} \;\; . \eqno(3.10)$$

In order to solve the equations of motion (3.8) we
introduce the double generating function of $f_{k,m}(t)$,
over the time variable $t$ and over the number of charges
$m$, with fixed number of sites $k$,

$$ g_k (u,w ) = \sum_{t=0}^\infty \sum_{m=0}^\infty
f_{k,m}(t) u^t w^m \;\; . \eqno(3.11) $$

\noindent It is also convenient to introduce the variable
$a=r-\ell$ directly measuring the hopping anisotropy,

$$ r=(1+a)/2 \;\;\;\;\;\; {\rm and} \;\;\;\;\;\;
\ell=(1-a)/2 \;\; . \eqno(3.12) $$

\noindent A straightforward but tedious calculation then
yields the following difference equation which derives from
the equations of motion (3.8), with(3.9),

$$g_{k+1}(u,w)+2{\left(1+a^2\right)u-2 \over
\left(1-a^2\right)u}g_k(u,w) +g_{k-1}(u,w)=-{4 \over
\left(1-a^2\right)u}(w p+1-p)^k \;\; . \eqno(3.13)$$

\noindent The initial and boundary conditions ``translate''
as follows,

$$g_k(0,w)=(w p +1-p)^k \;\; , \eqno(3.14)$$

$$g_0(u,w)={1 \over 1-u} \;\; . \eqno(3.15)$$

The solution of (3.13) is obtained as a linear combination
of the special solution $ \Omega (w p+1-p)^k $ proportional
to the right-hand side, and that solution of the
homogeneous equation which is regular at $u=0$. The
coefficient $\Omega $ is obtained by substitution,

$$\Omega =-{4(w p+1-p) \over \left(1-a^2\right)u \left(w
p+1-p-\Lambda_+\right)\left(w p+1-p-\Lambda_-\right)} \;\; ,
\eqno(3.16)$$

\noindent where it is convenient to express the denominator
in terms of the roots $\Lambda_\pm$ of the characteristic
equation of (3.13),

$$\Lambda^2+2{\left(1+a^2\right)u-2 \over
\left(1-a^2\right)u}\Lambda+1=0 \;\; . \eqno(3.17)$$

\noindent These roots are given by

$$\Lambda_\pm={2-\left(1+a^2\right)u\pm
2\sqrt{(1-u)\left(1-a^2u\right)} \over \left(1-a^2\right)u}
\;\; , \eqno(3.18)$$

\noindent and the root $\Lambda_-$, which is nonsingular as
$u \to 0$, also gives the homogeneous solution proportional
to $ \Lambda_-^k$. The proportionality constant is
determined by (3.15).
In summary, the solution takes the form

$$g_k(u,w)=\left({1 \over 1-u}-\Omega
\right)\Lambda_-^k+\Omega (w p+1-p)^k \;\; . \eqno(3.19) $$

The $w$-dependence of this result is via $\Omega$, see
(3.16), and it is of a simple rational form. Therefore,
derivation of the dependence on the number of charges,
``generated'' by $w$, is relatively simple. In our
applications we will concentrate on densities of reactants
at single lattice sites, derived from $f_{k=1,m}(t)$. The
$m$-dependence then follows by expanding (3.19) in powers
of $w$. However, the $u$-dependence of (3.19) with (3.16),
(3.18), is more complicated. Therefore we will use the
generating functions for the time-dependence and most of
our explicit time-dependent expressions will be derived as
asymptotic results valid for large times. Indeed, the power
series in $u$ are then controlled by the singularity at
$u=1$, and analytical results can be derived by appropriate
expansions. Specifically, let us introduce the
time-generating function for the quantities $f_{1,m}(t)$
which represent the probability to find charge $m$ at a
lattice site at time $t$. We define

$$ G_m (u) = \sum_{t=0}^\infty f_{1,m}(t) u^t \;\; .
\eqno(3.20) $$

\noindent The central result of this section is thus

$$G_m(u)=\delta_{m,0}\left[{\Lambda_- \over 1-u}-{4 \over
\left(1-a^2\right)u} \right]-(-1)^m {4\Lambda_+p^m \over
\left(1-a^2\right)u\left(1-p-\Lambda_+\right)^{m+1}} \;\;
,\eqno(3.21)$$

\noindent where $\Lambda_\pm$ were given in (3.18). Note
that $G_m (u)$ is just the $m^{\rm th}$ Taylor series
coefficient in $w$, of the function $g_1 (u,w)$.

\vfil\eject

\noindent {\bf IV. SINGLE-SPECIES REACTIONS}

\

In this section we map the coagulating-charge model onto
the two single-species reactions introduced in Section~II.
This approach follows recent work [4], although related
ideas have been used in earlier literature, for instance,
in [35]. Let us consider first the two-particle aggregation
reaction $A+A\to A$. In the coagulating-charge model we now
regard each ``charged'' site as occupied by an
$A$-particle, and each ``uncharged'' site as empty of
$A$-particles. Specifically, any charge $m>0$ represents an
$A$ particle. No charge at a site, $m=0$, corresponds to
absence of an $A$ particle. The dynamics of the coagulating
charges then maps onto the dynamics of the reaction $A+A\to
A$ on the same lattice, with the initial conditions of
placing particles $A$ randomly with probability $p$, or
leaving the lattice sites vacant with probability $1-p$, so
that the initial density of the $A$ particles is

$$ c(0)=p \;\; . \eqno(4.1) $$

We now observe that the quantity $f_{1,0}(t)$ gives the
density of empty sites in both models. Therefore, the
particle density (per lattice site), $c(t)$, in the
aggregation model, is given by

$$ c(t)=1-f_{1,0}(t) \;\; . \eqno(4.2)$$

\noindent The generating function is therefore easily
derivable from (3.21),

$$ E(u)= \sum_{t=0}^\infty c(t)u^t={1 \over 1-u}-G_0(u) ={1
-\Lambda_- \over 1-u}+ {4(1-p) \over
\left(1-a^2\right)u\left(1-p-\Lambda_+\right)} \;\; .
\eqno(4.3)$$

The function $E(u)$ is actually regular at $u=0$. Thus the
Taylor series is controlled by the singularity at $u=1$,
near which we have the leading-order term

$$ E(u) = {2 \over \sqrt{1-a^2}} \left[{1 \over
\sqrt{1-u}}+{\cal O}(1) \right] \;\; . \eqno(4.4)$$

\noindent This yields the leading-order large-time
behavior,

$$ c(t) \approx {2 \over \sqrt{\left(1-a^2\right)\pi t\,}}
\;\; . \eqno(4.5) $$

\noindent While we are not aware of other exact solutions
for this model with anisotropic hopping, the result (4.5)
is not surprising. Indeed, the leading-order large time
behavior is expected to be universal in that it does not
depend on the initial density $p$. Furthermore, the
diffusion constant ${\cal D}(a)= \left( 1- a^2 \right)
{\cal D} (0)$ decreases proportional to $1-a^2$ when the
anisotropy is introduced (single-particle diffusion is then
of course also accompanied by drift). Therefore, as a
function of ${\cal D}(a) t$, the result (4.5) also does not
depend on the anisotropy and in fact it is the same as
expressions found for other $A+A\to A$ reaction models,
with different detailed dynamical rules (with isotropic
hopping), e.g., [20].

Let us now turn to the two-particle annihilation model
$A+A\to {\rm inert}$. The appropriate mapping here is to
identify odd charges with particles $A$ and even charges
with empty sites. Indeed, the dynamics of the
coagulating-charge model is then mapped onto the reaction
$A+A\to {\rm inert}$. Thus, each lattice site with an odd
charge $q=1,3,5,\ldots$ will
be replaced by a site with one $A$ particle.
Each lattice site with an
even charge $q=0,2,4,\ldots$ is empty of
$A$-particles. The rules of charge coagulation then reduce
to the desired reaction. Furthermore, relation (4.1)
applies. However, the generating function $E(u)$ for
particle density is given by a different expression,

$$ E(u) = \sum_{j=0}^\infty G_{2j+1}(u)= {4\Lambda_+ p
\over \left(1-a^2\right)u
\left[\left(1-p-\Lambda_+\right)^2-p^2\right]} \;\; .
\eqno(4.6)$$

\noindent The large-time behavior is similar to the
aggregation reaction, with the universal expression which
only differs from (4.5) by a factor of\ 2,

$$ c(t) \approx {1 \over \sqrt{\left(1-a^2\right)\pi t\,}}
\;\; . \eqno(4.7) $$

While the large-time behavior of both models is
model-independent and otherwise universal as described
earlier, the finite-time results do depend on details of
the dynamical rules. For our particular choice of
synchronous dynamics on alternating sublattices (Figure~1),
there exists an exact mapping relating the
isotropic-hopping aggregation and annihilation reactions
[4]. This mapping was also found for the
anisotropic-hopping results obtained here. Specifically, we
find (by comparing their generating functions)

$$ 2c_{\rm inert} (t;p)=c_A (t;2p) \;\; , \eqno(4.8) $$

\noindent where the subscripts denote the outcome of the
reaction while the added argument stands for the initial
density. Thus if we consider an annihilation reaction with,
initially, at $t=0$, half the particle density as compared
to an aggregation reaction with the same synchronous
dynamical rules, then the density ratio will remain exactly
1/2 for all later times, $t=1,2,\ldots$, as well.

We note that due to complexity of expressions involved,
results like (4.4) and especially (4.8) in this section as
well as many other expressions in the next two sections,
could only be derived by using symbolic computer programs
Maple and Mathematica.

\vfil\eject

\noindent {\bf V. TWO-SPECIES ANNIHILATION MODEL}

\

In this section we present results for the $AB$ model
defined in Section~II. We assume that initially particles
are placed with density $p$, but now a fraction $\alpha$ of
them are type $A$, and a fraction $\beta$ are type $B$.
Clearly,

$$\alpha+\beta=1 \;\; . \eqno(5.1) $$

\noindent Furthermore, the initial $A$- and $B$-particle
concentrations are, respectively, $\alpha p$ and $\beta p$.
The concentration difference is constant during the
reaction; it remains $(\alpha-\beta) p$. At large times,
this is also the limiting value of the density of the
majority species, while the density of the minority species
vanishes. In what follows we assume

$$\alpha \geq \beta \;\; , \eqno(5.2) $$

\noindent which can be done without loss of generality.
Indeed, the results for $\alpha < \beta$ can be obtained by
relabeling the particle species. Thus, either the
concentrations are equal or the majority species is always
$A$. Our goal will be to calculate the density, $c(t)$, of
the majority species $A$.

The dynamics of the $AB$ model can be related to that of
the coagulating-charge model of Section~III by adapting the
ideas of [28-29]. First, we note that the dynamics of the
``sticky'' $A+B \to {\rm inert}$ model can be viewed as
coagulation. Thus, if a group of $n$ particles $A$ and $m$
particles $B$ meet, they can be viewed as coagulating and
continuing to move together. Of course, at later times
this combined group may become part of a larger coagulated
cluster of particles. However, if their contribution to the
particle count is according to (2.2), the annihilation is
properly accounted for by this counting both upon original
and later coagulation events with other clusters.
Alternatively, one can view these particles as new charges,
$+1$ for $A$, and $-1$ for $B$. If the total charge of a
coagulated cluster is positive than we view it as a group
of $A$ particles (equal in their number to the charge
value). If the charge is negative, we consider the cluster
$B$-particle, while if the charge is $0$, we regard this
cluster as nonexistent (inert) in the $AB$ model.

The probability of having an $m$-particle (charge $m$)
cluster in the original positive-charge-only model was
given by $f_{1,m}(t)$. Each such $m$-particle cluster can
have charge $n = -m, -m+2, \dots, m-2, m$, where we now refer
to the new, $\pm$ charge definition rather than to the
positive charges of the original coagulation model. The key
observation is that having a ``species'' label assigned to
a particle at time $t=0$ is statistically independent of
its motion and affiliation as part of clusters at later
times. Statistical averaging over the particle placement
initially, and their motion at later times while coagulating
to form particle clusters, is uncorrelated with the choice,
with probabilities $\alpha$ and $\beta$, of which species
label to allocate to each particle at time $t=0$.

Thus the density (per site) of $m$-size clusters with
exactly $n$ units of charge, where $m$-size is that of the
original coagulation models, while the charge $-m \leq n
\leq m$ is the new $\pm$ type, is given by

$$\Psi_{m,n}(t)={\alpha\vphantom{T}}^{m+n \over
2}{\beta\vphantom{T}}^{m-n \over 2} {m! \over \left({ m+n
\over 2}\right)!\, \left({ m-n \over 2}\right)!} f_{1,m}(t)
\;\; . \eqno(5.3)$$

\noindent Therefore the density per site of $A$-particles,
i.e., the density per site of the $+$ charge, can be
written as

$$ c(t)=\sum_{n=1}^\infty n \left[
\sum_{m=n,n+2,\ldots}\Psi_{m,n}(t) \right] \;\; .
\eqno(5.4) $$

Similar to Section~IV, let us denote the time-generation
function of $c(t)$ by $E(u)$; see (4.3). By using
results from Section~III, and after some algebra, we get
the following expression for the generating function,

$$ E(u)={4\Lambda_+ \over
\left(1-a^2\right)u\left(p+\Lambda_+ -1 \right)}
\left(x{\partial \over \partial x} -y{\partial \over
\partial y}\right) S(x,y) \;\; . \eqno(5.5) $$

\noindent Here we introduced the function

$$ S(x,y)= \sum_{n=1}^\infty \, \sum_{j=0}^\infty \,
x^{n+j} y^{j}{n+2j \choose j} \;\; , \eqno(5.6)$$

\noindent and the variables

$$ x={p\alpha \over p+\Lambda_+ -1} \;\; , \eqno(5.7) $$

$$ y={p\beta \over p+\Lambda_+ -1} \;\; . \eqno(5.8) $$

To make progress, we have to evaluate the double-sum in
(5.6). This is accomplished as follows. First, we write
the sum

$$S(x,y)=\sum_{n=1}^{\infty}s_nx^n \;\; , \eqno (5.9) $$

\noindent where

$$s_n=\sum_{j=0}^{\infty} {(xy)^j \, \Gamma(2j+n+1) \over
j! \, \Gamma(j+n+1)} \;\; . \eqno(5.10)$$

\noindent Now, by using the duplication formula for the
gamma function, we have

$$s_n=\sum_{j=0}^{\infty} { \left({n+1 \over 2}\right)_j
\left({n+2 \over 2}\right)_j (4xy)^j \over j! \, (n+1)_j }
\;\; , \eqno(5.11) $$

\noindent where $(z)_j=\Gamma(z+j)/\Gamma(z)$. The sum in
$s_n$ is a special case of the hypergeometric function,

$$s_n=\;_2F_1(\nu,\nu+1/2;2\nu;\zeta) \;\; , \eqno(5.12)$$

\noindent where $\nu=(n+1)/2$ and $\zeta=4xy$.

Fortunately, this can be expressed in elementary terms; all
subsequent references in this paragraph are to formulas in
Chapter 15 of [36]. First, using Gauss' linear
transformation (15.3.6),

\

$ s_n={\Gamma(2\nu)\Gamma(-1/2)\over
\Gamma(\nu)\Gamma(\nu-1/2)}\;_2F_1(\nu,\nu+1/2;3/2;
1-\zeta)$\hfill

$\hbox{\hphantom{$s_n$}}+{\Gamma(2\nu)\Gamma(1/2)\over
\Gamma(\nu)\Gamma(\nu+1/2)}(1-\zeta)^{-1/2}\;_2F_1(
\nu,\nu-1/2;1/2;1-\zeta) \;\;$.\hfill (5.13)\break

\noindent Next, from the recursion relation (15.2.20),

\

\hang{}$\;_2F_1(\nu,\nu-1/2;1/2;1-\zeta)
=\zeta\;_2F_1(\nu,\nu+1/2;1/2;1-\zeta)$\hfill\break
$\hbox{\hphantom{$\;_2F_1(\nu,\nu-1/2;1/2;1-\zeta)$}} +
(1-2\nu)(1-\zeta)\;_2F_1(\nu,\nu+1/2;3/2;1-\zeta) \;\;$
. \hfill (5.14)\break

\noindent Now, by (15.1.10),

$$\;_2F_1(\nu,\nu+1/2;3/2;1-\zeta)={(1-\zeta)^{-1/2} \over
2(1-2\nu)} \left[\left(1+\sqrt{1-\zeta}\right)^{1-2\nu}
-\left(1-\sqrt{1-\zeta}\right)^{1-2\nu}\right] \;\; ,
\eqno(5.15)$$

\noindent while (15.1.9) yields the analytical expression,

$$\;_2F_1(\nu,\nu+1/2;1/2;1-\zeta)={1\over
2}\left[\left(1+\sqrt{1-\zeta}\right)^{-2\nu}+\left(1-\sqrt{1-\zeta}
\right)^{-2\nu}\right] \;\; , \eqno(5.16)$$

\noindent and so we have

\

\hang{}$\;_2F_1(\nu,\nu-1/2;1/2;1-\zeta)={1 \over 2} \zeta
\left[\left(1+\sqrt{1-\zeta}\right)^{-2\nu}+\left(1-\sqrt{1-\zeta}
\right)^{-2\nu}i\right]$\hfill\break
$\hbox{\hphantom{$\;_2F_1(\nu,\nu-1/2;1/2;1-\zeta)$}}+ {1
\over 2}(1-\zeta)^{1/2}\left[\left(1+\sqrt{1-\zeta}
\right)^{1-2\nu}-\left(1-\sqrt{1-\zeta}\right)^{1-2\nu}\right]
\;\;$.\hfill\break \hphantom{A}\hfill (5.17)\break

\noindent Putting all this together we find

$$s_n=2^{2\nu-1}(1-\zeta)^{-1/2}\left(1+\sqrt{1-\zeta}\right)^{1-2\nu}
=(1-4xy)^{-1/2}\left({2 \over 1+\sqrt{1-4xy}}\right)^n \;\;
, \eqno(5.18)$$

\noindent and finally,

$$S(x,y)=(1-4xy)^{-1/2}\sum_{n=1}^\infty \left({2x \over
1+\sqrt{1-4xy}}\right)^n= {2x \over \sqrt{1-4xy\,}
\left(1-2x+\sqrt{1-4xy}\right)} \;\; . \eqno(5.19) $$

It is useful to introduce the parameter $b=\alpha-\beta
\geq 0$ which measures the excess of $A$ over $B$ at time
$t=0$,

$$ \alpha=(1+b)/2 \;\;\;\;\;\; {\rm and} \;\;\;\;\;\;
\beta=(1-b)/2 \;\; . \eqno(5.20)$$

\noindent Consider first the equal concentration case
$b=0$. The large-time behavior of the concentration $c(t)$
is governed by the singularity at $u=1$ of the generating
function $E(u)$. The form of the latter was evaluated near
$u=1$ from the expressions derived in this section, with
the result,

$$E(u) = {1 \over (1-u)^{3/4}}\left[{\sqrt{p} \over
2\left(1-a^2\right)^{1/4}} -{1-p \over
4\sqrt{p}\left(1-a^2\right)^{3/4}}(1-u)^{1/2}+{\cal
O}(1-u)\right] \;\; . \eqno(5.21)$$

\noindent The leading-order behavior of the $A$-particle
concentration follows from the first term in (5.21), while
the second term will be further discussed in Section~VI. We
get

$$c (t) \approx {\sqrt{p} \over
2\Gamma(3/4)\left(1-a^2\right)^{1/4} t^{1/4}} \;\; .
\eqno(5.22)$$

\noindent The most significant feature of this result is
that, similar to the single-species reactions considered in
Section~IV, the anisotropy, $a$, dependence can be fully
absorbed in the diffusion constant, in terms of ${\cal
D}(a)t = \left(1-a^2\right) {\cal D}(0)t$. The exponent
$1/4$ was derived in [28-29] for different (isotropic)
dynamical rules.

A similar expansion for fixed $b>0$ yields

$$E(u) = {bp \over 1-u}+{1-b^2 \over
\left(1-a^2\right)b^3p}
-{2\left(1-b^2\right)\left(2-b^2p\right) \over
\left(1-a^2\right)^{3/2}b^5p^2} \sqrt{1-u\,}+{\cal O}(1-u)
\;\; . \eqno(5.23)$$

\noindent The leading term in (5.23) corresponds to the
constant contribution $c(t) = bp + \ldots$ which is
expected since $A$ is the majority species. In fact,
expansions near $u=1$ are nonuniform in the limits $b\to
0^+$ and $b\to 0^-$. In deriving (5.23) we used for the
first time the fact that the majority species is $A$. The
approach to the constant asymptotic density value is given
by the third term in (5.23),

$$c(t) -bp \approx {\left(1-b^2\right) \left(2-b^2p\right)
\over \sqrt{\pi} \, b^5 p^2 \left(1-a^2\right)^{3/2}
t^{3/2}} \;\; . \eqno(5.24)$$

\noindent Note that this difference is just the density of
the minority species $B$. As before, the anisotropy
dependence of this leading-order power-law correction is
fully absorbed in the diffusion rate, while the exponent is
consistent with the results of [28-29]. Details of the
crossover in the limit $b \to 0$ will be discussed in the
next section.

\vfil\eject

\noindent {\bf VI. CROSSOVER SCALING IN THE TWO-SPECIES
REACTION}

\

As emphasized in the preceding section, the limit $u \to 1$
is nonuniform at $b=0$, i.e., the pattern of the asymptotic
large-time behavior changes at equal $A$- and $B$-particle
concentrations. It is of interest to explore this behavior
in greater detail within the standard crossover scaling
formulation. In this approach, one seeks a combination of
powers of variables each of which vanishes in the limit of
interest, such that this so-called scaling combination can
be kept fixed in the double-limit. The appropriate choice
is expected to yield a nontrivial variation of quantities of
interest in the limit, as functions of the scaling
combination.

In our case, the appropriate scaling combination turns out
to be proportional to $b/(1-u)^{1/4}$, as determined by
inspection of various limiting expressions. It proves
convenient to absorb certain constants into the precise
definition of the scaling combination $\sigma$,

$$ \sigma = \sqrt{p}\,\left(1-a^2\right)^{1/4}b/(1-u)^{1/4}
\;\; . \eqno(6.1) $$

\noindent The time-generating function $E(u)$ studied in
Section~V will be now analyzed in the double-limit $b \to
0$ and $u \to 1^-$, taken with fixed values
of $\sigma$. From expressions derived in
Section~V, one can obtain

$$ E(u) \approx p^{-1} \left(1-a^2\right)^{-1} b^{-3}
R(\sigma) \;\; , \eqno(6.2) $$

\noindent where $R$ is termed the scaling function. Note
that the first two prefactors are constants in the limit of
interest. However, the power $b^{-3}$ is necessary to
ensure scaling function values of order 1 for $\sigma$ of
order 1.

The scaling function $R$ can be derived exactly,

$$ R(\sigma) = {\sigma^3
\left(\sigma+\sqrt{4+\sigma^2}\right)^2 \over 4
\sqrt{4+\sigma^2}} \;\; . \eqno(6.3)$$

\noindent Note that it is analytic at $\sigma=0$, where
$\sigma \propto b$. Thus, at the expense of introducing the
nonanalytic factor in $\sigma$ which is power-law in $1-u$,
see (6.1), we managed to ``blow up'' the regime of small
$b$. The scaling limit provides, as usual, a better
understanding of the crossover in the limit $b \to 0$. Note
that for $\sigma \ll 1$ the following small-argument
expansion of $R(\sigma)$ applies,

$$ R(\sigma)={1 \over 2}\sigma^3+{1 \over 2}\sigma^4+{\cal
O}\left(\sigma^5\right) \;\; . \eqno(6.4) $$

\noindent It is interesting to note that the leading term
here actually reproduces the first term in (5.21). The
latter was the limiting form for $u \to 1$ at $b=0$.
Indeed, the $b$-dependence cancels out, while the
$(1-u)$-dependence is the identical, simple power-law in both
limits. However, the second term in (5.21) does not seem to
correspond to the next scaling-expansion contribution; see
(6.4). Corrections to the leading scaling behavior
contribute to this term in the $b=0$ expansion.

In the opposite limit, $\sigma \to +\infty$, we get the
expansion

$$ R(\sigma)=\sigma^4+1-4 \sigma^{-2}+{\cal
O}\left(\sigma^{-4}\right) \;\; . \eqno(6.5)$$

\noindent The first term here reproduces the leading term
in (5.23). Indeed, the limit $\sigma \to +\infty$
corresponds to $u \to 1$ at fixed small positive $b$.
Interestingly enough, the next two terms in (5.23) are also
reproduced in their small-$b$ form by the next two terms in
(6.5). For instance, the second term in (6.5) yields
$1/\left[\left(1-a^2\right)b^3p\right]$ in $E(u)$.
Similarly, the third term in (5.23) is reproduced with
numerator 4 which is the correct small-$b$ limiting value.
Thus the minority-species concentration (5.24) with, for
small $b$, the numerator replaced again by 4, is also
contained in the scaling form. Of course, corrections to
scaling, not discussed here, yield improved results.

The main point of the scaling description is that it
provides a uniform limiting approximation in the
double-limit $b\to 0$ and $u\to 1$. Specifically, the
region of nonuniform behavior near $b=0$ is exploded by the
large factor $\sim (1-u)^{-1/4}$. In terms of $\sigma$, the
behavior is smooth and well defined. For instance, the
result (6.3) applies equally well for $\sigma<0$
which corresponds to $A$
becoming the minority species. The limit of $u\to 1^-$ at
small fixed $b<0$ is described by the limit $\sigma \to
-\infty$. The appropriate expansion takes the form

$$ R(\sigma)=-1+4 \sigma^{-2}+{\cal
O}\left(\sigma^{-4}\right) \;\; , \eqno(6.6)$$

\noindent similar in structure to (6.5) but without the
constant-density first term.

In summary, we derived exact results for several
reaction-diffusion models in $1D$. The leading-order
large-time particle densities show expected power-law and
universal behaviors. Anisotropy of hopping has no effect
on the universality class of the models studied, and it can
be largely absorbed into the definition of the diffusion
constant. While finite-time results are expected [26] to be
more sensitive to the value of the anisotropy parameter
$a$, they are cumbersome to derive and of less interest
than the leading-order expressions. One interesting
exception is the duality relation (4.8) which applies for
all finite time values in our synchronous-dynamics
models.

\vfil\eject

\noindent {\bf REFERENCES}

\

{\frenchspacing

\item{[1]} T. Liggett, {\sl Interacting Particle Systems\/}
(Springer-Verlag, New York, 1985).

\item{[2]} V. Kuzovkov and E. Kotomin, Rep. Prog. Phys.
{\bf 51}, 1479 (1988).

\item{[3]} V. Privman, in {\sl Trends in Statistical
Physics}, in print (Council for Scientific Information,
Trivandrum, India).

\item{[4]} V. Privman, Phys. Rev. E{\bf 50}, 50 (1994).

\item{[5]} H. Takayasu, Phys. Rev. Lett. {\bf 63}, 2563
(1989).

\item{[6]} D. Toussaint and F. Wilczek, J. Chem. Phys.
{\bf 78}, 2642 (1983).

\item{[7]} K. Kang and S. Redner, Phys. Rev. Lett. {\bf
52}, 955 (1984).

\item{[8]} K. Kang, P. Meakin, J.H. Oh and S. Redner, J.
Phys. A{\bf 17}, L665 (1984).

\item{[9]} K. Kang and S. Redner, Phys. Rev. A{\bf 32}, 435
(1985).

\item{[10]} S. Cornell, M. Droz and B. Chopard, Phys. Rev.
A{\bf 44}, 4826 (1991).

\item{[11]} V. Privman and M.D. Grynberg, J. Phys. A{\bf
25}, 6575 (1992).

\item{[12]} B.P. Lee, J. Phys. A{\bf 27}, 2533 (1994).

\item{[13]} M. Bramson and D. Griffeath, Ann. Prob. {\bf
8}, 183 (1980).

\item{[14]} D.C. Torney and H.M. McConnell, J. Phys. Chem.
{\bf 87}, 1941 (1983).

\item{[15]} Z. Racz, Phys. Rev. Lett. {\bf 55}, 1707
(1985).

\item{[16]} A.A. Lushnikov, Phys. Lett. A{\bf 120}, 135
(1987).

\item{[17]} M. Bramson and J.L. Lebowitz, Phys. Rev. Lett.
{\bf 61}, 2397 (1988).

\item{[18]} D.J. Balding and N.J.B. Green, Phys. Rev. A{\bf
40}, 4585 (1989).

\item{[19]} J.G. Amar and F. Family, Phys. Rev. A{\bf 41},
3258 (1990).

\item{[20]} D. ben-Avraham, M.A. Burschka and C.R. Doering,
J. Statist. Phys. {\bf 60}, 695 (1990).

\item{[21]} M. Bramson and J.L. Lebowitz, J. Statist.
Phys. {\bf 62}, 297 (1991).

\item{[22]} V. Privman, J. Statist. Phys. {\bf 69}, 629
(1992).

\item{[23]} R. Kopelman, C.S. Li and Z.--Y. Shi, J.
Luminescence {\bf 45}, 40 (1990).

\item{[24]} R. Kroon, H. Fleurent and R. Sprik, Phys. Rev.
E{\bf 47}, 2462 (1993).

\item{[25]} S.A. Janowsky, Phys. Rev. E, in print.

\item{[26]} V. Privman, J. Statist. Phys. {\bf 72}, 845
(1993).

\item{[27]} V. Privman, E. Burgos and M.D. Grynberg,
preprint.

\item{[28]} P. Krapivsky, Physica A{\bf 198}, 135 (1993).

\item{[29]} P. Krapivsky, Physica A{\bf 198}, 150 (1993).

\item{[30]} V. Privman, A.M.R. Cadilhe and M.L. Glasser, preprint.

\item{[31]} I.M. Sokolov and A. Blumen, Phys. Rev. E{\bf
50}, 2335 (1994).

\item{[32]} H. Takayasu, M. Takayasu, A. Provata
and G. Huber, J. Statist. Phys. {\bf 65}, 725 (1991).

\item{[33]} S.N. Majumdar and C. Sire, Phys. Rev. Lett.
{\bf 71}, 3729 (1993).

\item{[34]} H. Park, M. Ha and I.-M. Kim, preprint.

\item{[35]} J.L. Spouge, Phys. Rev. Lett. {\bf 60}, 871
(1988).

\item{[36]} M. Abramowitz and I.A. Stegun, {\sl Handbook of
Mathematical Functions}, (Dover, New York, 1972).

}

\vfil\eject

\noindent {\bf FIGURE CAPTIONS}

\

\noindent\hang{\sl Figure 1}:\ \ The $1D$ even-odd sublattices
represented as the two-dimensional space-time lattice directed along
the time axis. The solid-line bonds show possible
hopping event directions.

\

\noindent\hang{\sl Figure 2}:\ \ The charge in a continuous span of
$k$ lattice sites at time $t+1$ can come from (a) $k-1$ sites at
time $t$, with probability $r\ell$. Note that the directions of hopping
from these $k-1$ sites, shown by double-arrows, are immaterial. Only
the two exterior sites of the larger, $(k+1)$-interval shown
determine the
probability $r \ell$. Another possibility is (b) for the charge to come
from $k$ sites at time $t$. In this case both end-sites of the
$(k+1)$-interval hopped to the right. The probability of (b) is
therefore $r^2$. Similarly, the probability of the charge coming from
the other $k$-interval, event shown as (c), is $\ell^2$. Finally, (d)
the charge can also come from all $k+1$ sites shown at time $t$, with
probability $r\ell$.

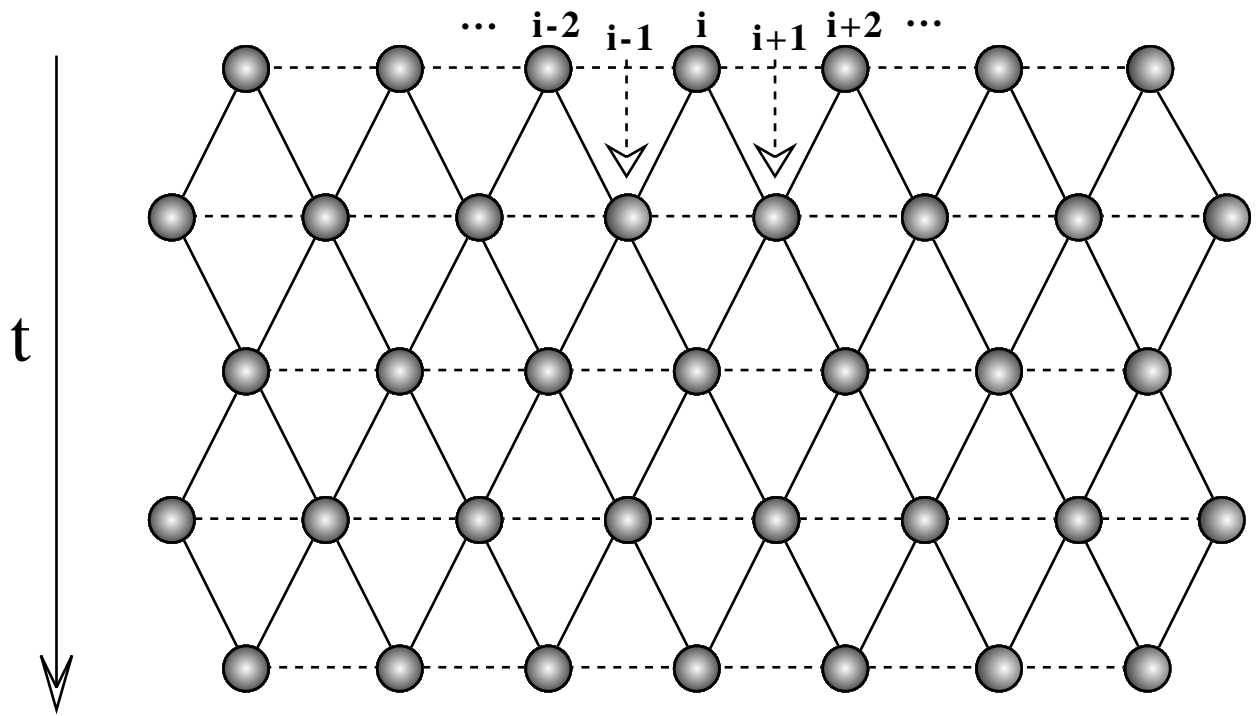
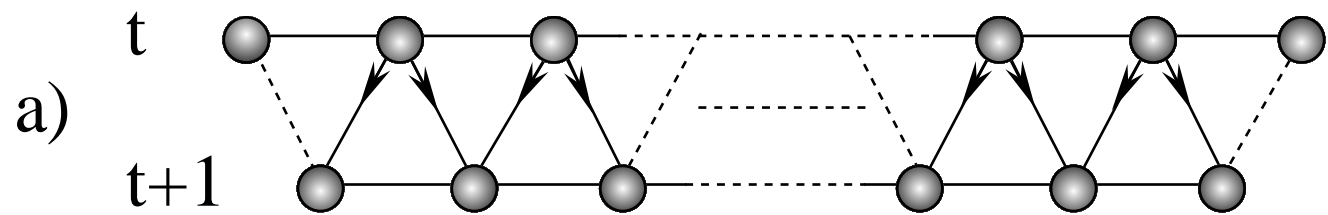
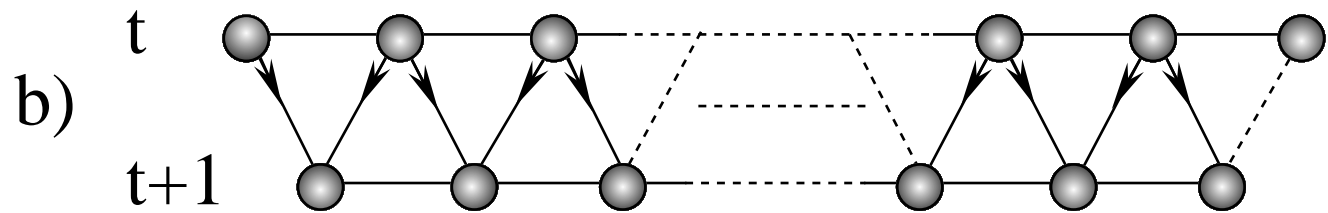
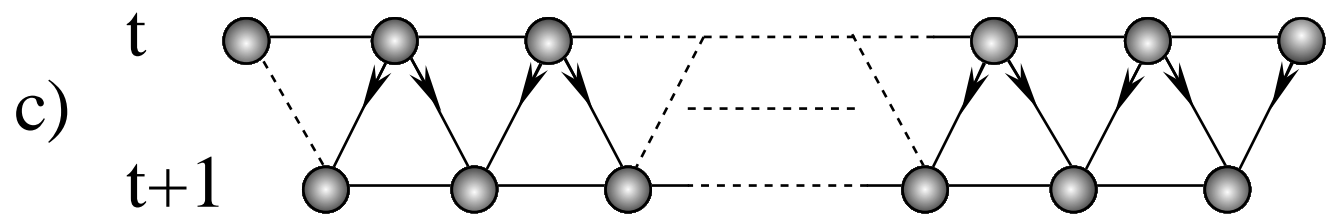
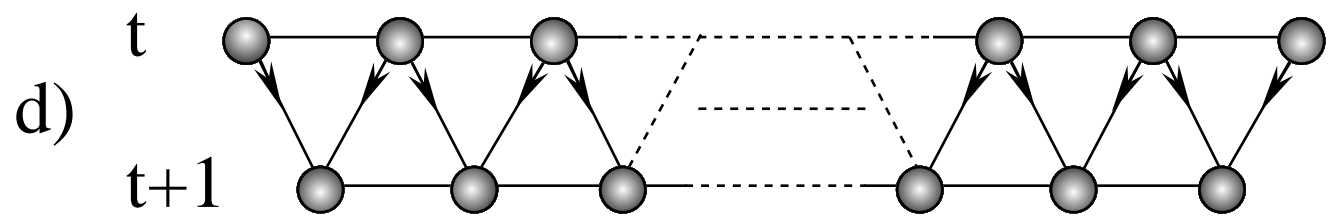
\bye